\documentclass[a4paper]{article}
\usepackage{geometry}
\usepackage{amsmath}
\usepackage{amsfonts}
\usepackage{amsmath,amssymb}
\usepackage{graphicx}
\usepackage{color}
\usepackage{bbm}
\usepackage{mathdots}

\newtheorem{theorem}{Theorem}
\newtheorem{lemma}{Lemma}
\newtheorem{corollary}{Corollary}

\newtheorem{definition}{Definition}

\newcommand{\ls}[1]
    {\dimen0=\fontdimen6\the\font\lineskip=#1\dimen0
     \advance\lineskip.5\fontdimen5\the\font
     \advance\lineskip-\dimen0
     \lineskiplimit=0.9\lineskip
     \baselineskip=\lineskip
     \advance\baselineskip\dimen0
     \normallineskip\lineskip\normallineskiplimit\lineskiplimit
     \normalbaselineskip\baselineskip
     \ignorespaces}

\begin{document}

\bibliographystyle{abbrv}

\title{Hermitian Self-dual Twisted Generalized Reed-Solomon Codes}
\author{Chun'e Zhao$^{1}$, Yuxin Han$^{1}$, Wenping Ma$^{2}$, Tongjiang Yan$^{1}$, Yuhua Sun$^{1}$\\
$^1$ College of Sciences,
China University of Petroleum,\\
Qingdao 266555,
Shandong, China\\
$^2$ School of Telecommunication Engineering,\\
 Xidian University,
  Xi'an, China\\
Email: zhaochune1981@163.com;\ wp\_ma@mail.xidian.edu.cn;\\
 yantoji@163.com;\ sunyuhua\_1@163.com\\
}
 \maketitle
\footnotetext[1] {.
}
\thispagestyle{plain} \setcounter{page}{1}
\begin{abstract}
Self-dual maximum distance separable (MDS) codes over finite fields are linear codes with significant combinatorial and cryptographic applications. Twisted generalized Reed-Solomon (TGRS) codes can be both MDS and self-dual. In this paper, we study a general class of TGRS codes (A-TGRS) codes, which encompasses all previously known special cases. First, we establish a sufficient and necessary condition for an A-TGRS code to be Hermitian self-dual. Furthermore, we present four constructions of self-dual TGRS codes, which, to the best of our knowledge, nearly cover all the related results previously reported in the literature. More importantly, we also obtain several new classes of Hermitian self-dual TGRS codes with flexible parameters. Based on this framework, we derive a sufficient and necessary condition for an A-TGRS code to be Hermitian self-dual and MDS. In addition, we construct a class of MDS Hermitian self-dual TGRS code by appropriately selecting the evaluation points. This work investigates the Hermitian self-duality of TGRS codes from the perspective of matrix representation, leading to more concise and transparent analysis. More generally, the Euclidean self-dual TGRS codes and the Hermitian self-dual GRS codes can also be understood easily from this point.

\end{abstract}

\ls{1.5}
\section{INTRODUCTION}\label{section 1}
Let et $\mathbb{F}_{q}$ be a finite filed of size $q$, where $q$ is a prime power and $\mathbb{F}_{q}^{*}=\mathbb{F}_{q}\backslash \{0\}$. An $[n, k, d]$ linear code $C$ is a subspace of the linear space $\mathbb{F}_ {q}^{n}$ over $\mathbb{F}_ {q}$ with dimension $k$ and minimum Hamming distance $d$. It is well known that the parameters of $C$ satisfy $d\leq n-k+1$. If $d=n-k+1$, $C$ is called maximum distance separable (MDS).

Generalized Reed-Solomon (GRS) codes are an important class of MDS linear codes. By adding some monomials (called twists) to different positions (called hooks) of each codeword polynomial, a GRS code can be generalized to twisted GRS (TGRS) code. However, a TGRS code is not necessarily MDS. After Beelen et al. firstly introduced the notion of TGRS code in 2017 \cite {8}, many researchers began to investigate the condition under which a TGRS code is MDS\cite{8,4,6,2,3,9,15,5,1,7,10-m,11-1,14,19,23,26,29}. The A-TGRS codes studied in \cite{29} were the most general form, then the study on the MDS conditions of TGRS codes is perfected. \\

Self-dual codes are another interesting class of codes due to their fascinating links to other objects and their wide applications in other fields such as lattices, cryptography, invariant theory, block designs, etc. Both Euclidean and Hermitian self-dual codes are also closely related to quantum stablizer codes. In recent years, there are also many studies on Euclidean self-duality of TGRS codes\cite{4,6,9,15,5,1,7,12-1,26}. However, there is only one article on the study of Hermitian self-dual TGRS codes \cite{12-1} and the code studied is also a very special type of TGRS code. For GRS codes which are also a special type of TGRS codes with 0 parameter matrix, there are two articles studying Hermitian self-duality\cite{37,38}. Besides these, the majority research on Hermitian self-dual is mainly about cyclic codes and constant cyclic codes, such as \cite{30,31,32,33,34,35,36}. Upon the above analysis, we choose to study the Hermitian self-duality of arbitrary TGRS (A-TGRS) codes in this paper.

In this paper, we transfer a TGRS code's Hermitian duality to another code's Euclidean duality and use their generator matrices instead of the traditional polynomial methods to study the problem of codeword equality, which greatly simplified the complexity of expressions. Based on this, we first give the sufficient and necessary condition under which an A-TGRS code is Hermitian self-dual. Through analysing the matrices in the expression, we find that when the evaluation points $\{\alpha_{1},\alpha_{2},\cdots,\alpha_{n}\}$ and the parameter matrix $A(\eta)$ are suitably selected, the corresponding TGRS codes can be Hermitian self-dual. Correspondingly, we give four constructions for Hermitian self-dual TGRS codes which can generalize the known constructions of Hermitian self-dual TGRS codes and also can generate many new classes of Hermitian self-dual TGRS codes. More importantly, we also give a sufficient and necessary condition under which an A-TGRS code is Hermitian self-dual and  MDS. Furthermore, we also give an explicit construction for Hermitian self-dual MDS TGRS codes. In addition,  when the parameters are limited in $\mathbb{F}_{q}$, the Hermitian self-duality can degenerate into Euclidean self-duality and we also give the sufficient and necessary condition under which an A-TGRS code is Euclidean self-dual and all the existing Euclidean self-dual TGRS codes can be considered as special cases. When the parameter matrix $A({\eta})=0$, the Hermitian self-dual TGRS codes can be degenerated into Hermitian self-dual GRS codes and the existing Hermitian self-dual GRS codes can also be viewed as special ones.

This paper is organized as follows. In Section $\mathrm{\uppercase\expandafter{\romannumeral2}}$, we recall some basic notations and symbols about A-TGRS codes. In section $\mathrm{\uppercase\expandafter{\romannumeral3}}$, we give the sufficient and necessary conditions under which the A-TGRS codes are Hermitian self-dual. In Section $\mathrm{\uppercase\expandafter{\romannumeral4}}$, we present four constructions for Hermitian self-dual TGRS codes. In Section $\mathrm{\uppercase\expandafter{\romannumeral5}}$, we give the sufficient and necessary conditions for an A-TGRS code to be MDS and self-dual and we also give an explicit construction.  Finally, we conclude our work in Section $\mathrm{\uppercase\expandafter{\romannumeral6}}$.

\section{Preliminaries}\label{section 2}

Let $q$ be a prime power and $\mathbb{F}_{q}$ be the finite field with $q$ elements. $\mathbb{F}_{q}^{*}=\mathbb{F}_{q}\backslash \{0\}$. Let $\mathbb{F}_{q}^{n}$ denote the vector space of all $n-$tuples over the finite field $\mathbb{F}_{q}$. If $C$ is a $k-$dimensional subspace of $\mathbb{F}_{q}^{n}$, then $C$ will be called an $[n,k]$ linear code over $\mathbb{F}_{q}$. The linear code $C$ has $q^{k}$ codewords.\\

Let ${ x}=(x_{1},x_{2},\cdots,x_{n}), { y}=(y_{1},y_{2},\cdots,y_{n})\in \mathbb{F}_{q}^{n}$, here we review that the Euclidean inner product of vectors ${ x},{ y}$ is
$$<{ x},{ y}>_{E}=\sum\limits_{i=1}^{n}x_{i}y_{i}.$$
The Euclidean dual code of $C$ is defined as

$$C^{\bot  E}=\{{ x}|{ x}\in \mathbb{F}_{q}^{n},<{ x},{ y}>_{E}=0, for\  all \ { y} \in C\}.$$

It is always useful to consider another inner product, called the Hermitian inner product.

$$<{ x},{ y}>_{H}=\sum\limits_{i=1}^{n}x_{i}y_{i}^{q}.$$

Similarly to the Euclidean case, we can define the Hermitian dual of $C$ by using this inner product as follows.
$$C^{\bot H}=\{{ x}|{ x}\in \mathbb{F}_{q}^{n},<{ x},{ y}>_{H}=0, for\ all\ { y}\in C\}.$$

Namely, $C^{\bot H}$ is the orthogonal subspace to $C$, with respect to the Hermitian inner product. We also have Hermitian self-orthogonality and Hermitian self-duality. If $C\subseteq C^{\bot H}$, then $C$ is Hermitian self-orthogonal. Particularly, if $C^{\bot H}=C$, then $C$ is Hermitian self-dual.

\subsection{TGRS codes and A-TGRS codes}
Let $\mathbb{F}_{q}[x]$ be the polynomial ring over $\mathbb{F}_{q}$. Denote $\mathbb{F}_{q}[x]_{n}=\{a_{n-1}x^{n-1}+\cdots+a_{1}x+a_{0}|a_{i}\in \mathbb{F}_{q},\  i=0,1,\cdots,n-1\}$. Suppose that ${\alpha}=(\alpha_{1},\alpha_{2},\cdots,\alpha_{n})\in\mathbb{F}_{q}^{n}$ and ${ v}=(v_{1},v_{2},\cdots,v_{n})\in\left(\mathbb{F}_{q}^{*}\right)^{n}$, where $\alpha_{1},\alpha_{2},\cdots,\alpha_{n}$ are distinct elements.

An $[n,k]$ generalized Reed-Solomon code $GRS_{n,k}({\alpha},{ v})$ associated with ${\alpha}$ and ${ v}$ is defined as
  $$GRS_{n,k}({\alpha},{ v}):=\{(v_{1}f(\alpha_{1}),v_{2}f(\alpha_{2}),\cdots,v_{n}f(\alpha_{n}))|f(x)\in \mathbb{F}_{q}[x]_{k}\}.$$
  After adding some polynomials into different positions of each $f(x)$ in $\mathbb{F}_{q}[x]_{k}$, the GRS code can be generalized as follows:

 \begin{definition}\cite{23}
 For positive integers $l,k$ and $n$ with $k<n$, let ${\eta}=(\eta_{1},\eta_{2},\cdots,\eta_{l})\in\left(\mathbb{F}_{q}^{*}\right)^{l}$, ${ h}=(h_{1},h_{2},\cdots,h_{l})\in\{0,1,\cdots,k-1\}^{l}$ and ${ t}=(t_{1},t_{2},\cdots,t_{l})\in \{1,2,\cdots,n-k\}^{l}$, where $(h_{i},t_{i})\neq (h_{j},t_{j})$ for all $1\leq i<j\leq l$. The set of $V_{(n,k,{ t},{ h},{\eta})}$ twisted polynomials are defined as
 $$ V_{(n,k,{ t},{ h},{\eta})}=\{f(x)=\sum\limits_{i=0}^{k-1}f_{i}x^{i}+\sum\limits_{j=1}^{l}\eta_{j}f_{h_{j}}x^{k-1+t_{j}}|a_{i}\in F_{q}\}.$$
 \end{definition}
\begin{definition}\cite{23}
Let ${ t},{ h},{\eta}$ and $V_{n,k,{ t},{ h},{\eta}}$ be defined as above. Let ${\alpha}=(a_{1},a_{2},\cdots,a_{n})\in \mathbb{F}_{q}^{n}$ with $a_{i}\neq a_{j}(i\neq j)$ and $\mathbf{v}=(v_{1},v_{2},\cdots,v_{n})\in ({\mathbb{F}_{q}}^{*})^{n}$. Then the TGRS code is defined as
$C_{n,k}({ t},{ h},{\eta},{\alpha},\mathbf{v})=\{v_{1}f(\alpha_{1}),v_{2}f(\alpha_{2}),\cdots,v_{n}f(\alpha_{n})\},$
where ${ h}$ is called the hook  vector and ${ t}$ is called the twist vector.
\end{definition}
  \begin{definition}\cite{23}
For a positive integer $ k\leq n\leq q$ and a $k\times(n-k)$ matrix
$$A({ \eta})=\left(\begin{array}{cccc}
 \eta_{0,1}&\eta_{0,2}& \cdots& \eta_{0,n-k}\\
\eta_{1,1} &  \eta_{1,2}&\cdots & \eta_{1,n-k} \\
\vdots& \vdots&\ddots & \vdots\\
 \eta_{k-1,1}& \eta_{k-1,2}&\cdots & \eta_{k-1,n-k}
 \end{array}
 \right)$$
over $\mathbb{F}_{q}$, the polynomial set
$$S=\left\{\sum\limits_{i=0}^{k-1}f_{i}x^{i}+\sum\limits_{i=0}^{k-1}f_{i}\sum\limits_{j=1}^{n-k}\eta_{i,j}x^{k-1+j}: \mathrm{for \ all\ } f_{i}\in \mathbb{F}_{q},0\leq i\leq k-1 \right\}$$
is a $k$-dimensional subspace of $\mathbb{F}_{q}[x]$. The linear code
$$TGRS_{n,k}({ \alpha},{ v},A({ \eta}))=\left\{v_{1}f(\alpha_{1}),v_{2}f(\alpha_{2}),\cdots,v_{n}f(\alpha_{n})|f(x)\in S\right\}$$
is called an arbitrary twisted generalized Reed-Solomon (A-TGRS) code.
\end{definition}
In fact, the code $TGRS_{n,k}({\alpha},{ v},A({ \eta}))$ has the following generator matrix:
\begin{equation}\label{generator matrix 1}
G=[I_{k},A({ \eta})]V_{n}({ \alpha})\mathrm{diag}({ v}),
\end{equation}
 where
$$I_{k}=\left(\begin{array}{cccc}
 1& & & \\
 &1& &  \\
 & &\ddots& \\
 & & & 1
 \end{array}
 \right),\ A({ \eta})=\left(\begin{array}{cccc}
 \eta_{0,1}&\eta_{0,2}& \cdots& \eta_{0,n-k}\\
 \eta_{1,1} &  \eta_{1,2}&\cdots & \eta_{1,n-k} \\
 \vdots& \vdots&\textcolor{blue}{\ddots} & \vdots\\
 \eta_{k-1,1}& \eta_{k-1,2}&\cdots & \eta_{k-1,n-k}
 \end{array}
 \right),\ V_{n}({ \alpha})=\left(\begin{array}{cccc}
1&1&\cdots&1\\
\alpha_{1}&\alpha_{2}&\cdots&\alpha_{n}\\
\vdots&\vdots&\textcolor{blue}{\ddots}&\vdots\\
\alpha_{1}^{n-1}&\alpha_{2}^{n-1}&\cdots&\alpha_{n}^{n-1}
\end{array}\right),$$ $\mathrm{diag}({ v})=\left(\begin{array}{cccc}
v_{1}&&&\\
&v_{2}&&\\
&&\ddots&\\
&&&v_{n}
\end{array}\right).$ The TGRS code generated by $G=[I_{k},A({ \eta})]V_{n}({ \alpha})\mathrm{diag}({ v})$ is denoted by $TGRS_{n,k}({ \alpha},{ v},A({ \eta}))$.\\

\section{Conditions for Hermitian Self-duality}

\begin{lemma}\label{C and Cq}
Let $C$ be an $[n,k]$ linear code over $\mathbb{F}_{q^{2}}$ and $C^{q}=\{(a_{1}^{q},a_{2}^{q},\cdots,a_{n}^{q})|(a_{1},a_{2},\cdots,a_{n})\in C\}$, then $C^{q}$ is also an $[n,k]$ linear code.
\end{lemma}

$\mathbf{Proof.}$
\begin{itemize}
\item[(1)] For every two vectors ${ x},{ y}\in C$, every $k\in \mathbb{F}_{q^{2}}$, it is obvious that ${ x}^{q}+{ y}^{q}=({ x}+{ y})^{q}\in C^{q}$ and $k{ x}^{q}=(k^{q}{ x})^{q}\in C^{q}$.

So $C^{q}$ is a linear subspace of $\mathbb{F}_{q^{2}}^{n}$.\\

\item[(2)]  Consider the map
 $$f: \begin{array}{ccc}
 C&\rightarrow &C^{q}\\
 (x_{1},x_{2},\cdots,x_{n})&\mapsto&(x_{1}^{q},x_{2}^{q},\cdots,x_{n}^{q})
 \end{array}.$$
  It is obvious that $f$ is bijective. So the cardinality $|C|=|C^{q}|$. Thus the dimensions satisfy $dim C^{q}=dimC=k$. Then $C^{q}$ is also an $[n,k]$ linear code.
 \end{itemize}
 \begin{lemma}\label{Hermitian self-dual dimention}
 Let $C$ be an $[n,k]$ linear code over $\mathbb{F}_{q^{2}}$, where $q$ is a prime power.  Then $C$ is Hermitian self-dual if and only if $n=2k$ and $C^{q}\subset C^{\bot E}$.
 \end{lemma}

$\mathbf{Proof.}$

$\Rightarrow$\\

If $C$ is Hermitian self-dual, then $C=C^{\bot H}$ and $dim C=dim C^{\bot H}$. By the definition of Hermitian self-dual, we know that $C^{\bot H}=(C^{q})^{\bot E}$. By Lemma \ref{C and Cq}, $dim C^{\bot H}=dim (C^{q})^{\bot E}=n-dim C^{q}=n-k$. So $dim C=dim C^{\bot H}=n-k$, i.e. $n=2k$. By the reason that $C=C^{\bot H}=(C^{q})^{\bot E}$, we have $C^{\bot E}=C^{q}$, then $C^{q}\subseteq C^{\bot E}$.\\

$\Leftarrow$\\

If $n=2k$, i.e. $k=n-k$. So $dim C^{\bot E}=dim (C^{q})$ holds. And $C^{q}\subseteq C^{\bot E}$ also holds, so $C^{q}= C^{\bot E}$, i.e. $C=(C^{q})^{\bot E}=C^{\bot H}$. So $C$ is Hermitian self-dual.

\begin{lemma}\label{parity check matrix}(Theorem 4.2 in \cite{1})
Let $\alpha_{1},\alpha_{2},\cdots,\alpha_{n}$ be chosen as $n$ distinct elements in $\mathbb{F}_{q}$ and $G(x)=\prod\limits_{i=1}^{n}(x-\alpha_{i})=\sum\limits_{j=0}^{n}c_{j}x^{n-j}$. Assume that ${\mathbf u}=(u_{1},u_{2},\cdots,u_{n})$ where
$u_{i}=\dfrac{1}{G'(\alpha_{i})}$ for $1\leq i\leq n$. Then for any ${\mathbf v}=(v_{1},v_{2},\cdots,v_{n})\in (\mathbb{F}_{q}^{*})^{n}$, the code $TGRS_{n,k}({\mathbf \alpha},{\mathbf v}, A({\mathbf\eta}))$ has a parity check matrix
$$H=[I_{n-k},-J_{n-k}A({\mathbf\eta})^{T}J_{k}]T({\mathbf \alpha})V_{n}({\mathbf \alpha})\mathrm{diag}({\mathbf u}/{\mathbf v}),$$
where ${\mathbf u}/{\mathbf v}=(u_{1}/v_{1},u_{2}/v_{2},\cdots,u_{n}/v_{n})$ and matrices $J_{k}$, $T({\mathbf \alpha})$ and $V_{n}({\mathbf \alpha})$ are defined as follows:
\begin{equation}\label{matix symbols 1}
J_{k}=\left(\begin{array}{cccc}
0&\cdots&0&1\\
0&\cdots&1&0\\
\vdots&\iddots&\vdots&\vdots\\
1&\cdots&0&0
\end{array}
\right), T({\mathbf \alpha})=\left(\begin{array}{cccc}
c_{0}&&&\\
c_{1}&c_{0}&&\\
\vdots&\vdots&\ddots&\\
c_{n-1}&c_{n-2}&\cdots&c_{0}
\end{array}
\right),
V_{n}({\mathbf \alpha})=\left(\begin{array}{cccc}
1&1&\cdots&1\\
\alpha_{1}&\alpha_{2}&\cdots&\alpha_{n}\\
\vdots&\vdots&\ddots&\vdots\\
\alpha_{1}^{n-1}&\alpha_{2}^{n-1}&\cdots&\alpha_{n}^{n-1}
\end{array}
\right).
\end{equation}
\end{lemma}

From the above analysis, we can get the following necessary and sufficient conditions under which the TGRS codes are Hermitian self-dual.

\begin{theorem}\label{sufficient and necessary condtions for Hermitian self-dual matrix}
Let $\mathbb{F}_{q^{2}}$ be a finite field with $q^{2}$ elements and $q$ be a prime power, $\alpha_{1},\alpha_{2},\cdots,\alpha_{n}$ be different elements in $\mathbb{F}_{q^{2}}$. Let $n=2k$, $A({\mathbf \eta})\in \mathbb{F}_{q^{2}}^{k\times k}$ be a parameter matrix, then the code $TGRS_{n,k}({\mathbf \alpha},{\mathbf v}, A({\mathbf \eta}))$ is Hermitian self-dual if and only if there exists an inverse matrix $P$ such that
\begin{equation}\label{Hermitian self-dual general}
[I_{k},A({\mathbf \eta})^{q}]V_{n}({\mathbf \alpha})^{q}\mathrm{diag}(\dfrac{{\mathbf v}^{q+1}}{\mathbf u})=P[I_{k},-JA({\mathbf \eta})^{T}J]T({\mathbf \alpha})V_{n}({\mathbf \alpha}),
\end{equation}
where $J, T({\mathbf \alpha}), V_{n}({\mathbf \alpha})$ are denoted as in Eq. (\ref{matix symbols 1}) and $A^{q}$ be the $q-$ power of every entry of matrix $A$.
\end{theorem}
$\mathbf{Proof.}$
Let $C$ be the TGRS code $TGRS_{n,k}({\mathbf \alpha},{\mathbf v}, A({\mathbf \eta}))$, and $n=2k$. Then by Lemma \ref{Hermitian self-dual dimention}, $C$ is Hermitian self-dual if and only if $C^{q}=C^{\bot E}$. Let $G=[I_{k},A({\mathbf \eta})]V_{n}({\mathbf \alpha})\mathrm{diag}({\mathbf v})$ and by Lemma \ref{C and Cq}, then $C^{q}$ is also a linear space over $\mathbb{F}_{q^{2}}$ with
$$G^{q}=[I_{k},A({\mathbf \eta})^{q}]V_{n}({\mathbf \alpha})^{q}\mathrm{diag}({\mathbf v}^{q})$$
as its generator matrix. By Lemma \ref{parity check matrix}, the parity matrix $H=[I_{k},-JA({\mathbf \eta})^{T}J]T({\mathbf \alpha})V_{n}({\mathbf \alpha})\mathrm{diag}({\mathbf u}/{\mathbf v})$ is the generator matrix of $C^{\bot E}$. So $C$ is Hermitian self-dual if and only if there exists an inverse matrix $P$, such that $[I_{k},A({\mathbf \eta})^{q}]V_{n}({\mathbf \alpha})^{q}\mathrm{diag}({\mathbf v}^{q})=P[I_{k},-JA({\mathbf \eta})^{T}J]T({\mathbf \alpha})V_{n}({\mathbf \alpha})\mathrm{diag}({\mathbf u}/{\mathbf v})$, which is equivalent to
$$[I_{k},A({\mathbf \eta})^{q}]V_{n}({\mathbf \alpha})^{q}\mathrm{diag}(\dfrac{{\mathbf v}^{q+1}}{\mathbf u})=P[I_{k},-JA({\mathbf \eta})^{T}J]T({\mathbf \alpha})V_{n}({\mathbf \alpha}).$$

When the parameters of $TGRS_{n,k}({\mathbf \alpha},{\mathbf v}, A({\mathbf \eta}))$ are limited to $\mathbb{F}_{q}$, the Hermitian self-dual code degenerates into Euclidean self-dual code. Then the following Corollary can be obtained immediately.

\begin{corollary}\label{Euclidean self-dual}
Let $\mathbb{F}_{q}$ be a finite field with $q$ elements and $q$ be a prime power, $\alpha_{1},\alpha_{2},\cdots,\alpha_{n}$ be different elements in $\mathbb{F}_{q}$. Let $n=2k$, $A({\mathbf \eta})\in \mathbb{F}_{q}^{k\times k}$ be a parameter matrix. Then $TGRS_{n,k}({\mathbf \alpha},{\mathbf v}, A({\mathbf \eta}))$ is Euclidean self-dual if and only if there exists an inverse matrix $P$ such that
$$[I_{k},A({\mathbf \eta})]V_{n}({\mathbf \alpha})\mathrm{diag}(\dfrac{{\mathbf v}^{2}}{\mathbf u})=P[I_{k},-JA({\mathbf \eta})^{T}J]T({\mathbf \alpha})V_{n}({\mathbf \alpha}).$$
\end{corollary}

{\bf Remark 1:} When $\frac{v_{i}^{2}}{u_{i}}=\lambda\in \mathbb{F}_{q}^{*}$, for $i=1,2,\cdots,n$, then $TGRS_{n,k}({\mathbf \alpha},{\mathbf v}, A({\mathbf \eta}))$ is Euclidean self-dual if and only if  there exists an inverse matrix $P$ such that
$$[I_{k},A({\mathbf \eta})]=P[I_{k},-JA({\mathbf \eta})^{T}J]T({\mathbf \alpha}).$$ So, the Euclidean self-dual TGRS codes constructed in \cite{6,9,15,5,1,7,12-1,26} can be understood more clearly and easily from this perspective.

\section{Constructions of Hermitian self-dual TGRS codes}

From Eq. (\ref{Hermitian self-dual general}), we can see that if the evaluation points $\alpha_{i}$'s satisfy $\alpha_{i}^{q}=a\alpha_{i}+b$ for some fixed $a,b\in \mathbb{F}_{q^{2}}^{*}$, for $i=1,2,\cdots,n$, then $V_{n}({\mathbf \alpha})^{q}$ will have a linear relationship with $V_{n}({\mathbf \alpha})$. Following, we will consider the roots of $x^{q}-ax-b=0$ in $\mathbb{F}_{q^{2}}$.

\begin{theorem}
Let $q$ be a prime power, $\mathbb{F}_{q^{2}}$ be a finite field with $q^{2}$ elements, $a,b \in\mathbb{F}_{q^{2}}$. Define a set
$$S=\{\alpha\in \mathbb{F}_{q^{2}}|\alpha^{q}=a\alpha+b\}.$$
Then the cardinality $|S|>1$ if and only if $a^{q+1}=1, b^{q}+a^{q}b=0$.
\end{theorem}
$\mathbf{Proof.}$

\begin{itemize}
\item[(1)] When $a=0,b=0$, it is obvious that $S=\{0\}$ and $|S|=1$.
\item[(2)] When $b=0,a\neq0$, $\alpha^{q}=a\alpha+b$ turns into $\alpha(\alpha^{q-1}-a)=0$. So $|S|>1$ if and only if $a^{q+1}=1$.
\item[(3)] When $b\neq0$, then by $\alpha^{q}=a\alpha+b$ we know that $\alpha=(a\alpha+b)^{q}=a^{q}\alpha^{q}+b^{q}=a^{q}(a\alpha+b)+b^{q}=a^{q+1}\alpha+a^{q}b+b^{q}$.
So if $a^{q+1}\neq1$, then $|S|=1$. And if $a^{q+1}=1$, then $|S|>1$ if and only if $a^{q}b+b^{q}=0$.
\end{itemize}
From (1)(2)(3), it indicates that $|S|>1$ if and only if $a^{q+1}=1, b^{q}+a^{q}b=0$.

Moreover, $S=\{\alpha\in \mathbb{F}_{q^{2}}|\alpha^{q}=a\alpha+b\}$ also has certain algebraic structure when $a^{q+1}=1, b^{q}+a^{q}b=0$.

\begin{theorem}\label{Hermitian patition}
Let $q$ be a prime power, $\mathbb{F}_{q^{2}}$ be a finite field with $q^{2}$ elements, $a\in\mathbb{F}_{q^{2}}$ satisfy $a^{q+1}=1$. Let $A_{(a,q)}=\{b|b^{q}+a^{q}b=0\}$, $U_{(a,q)}^{b}$ be the set of roots of the equation $x^{q}-ax-b=0$. Then
 \begin{itemize}
\item[(1)]  $U_{(a,q)}^{b}\subseteq \mathbb{F}_{q^{2}}$ for any $b\in A_{(a,q)}.$\\
\item[(2)]  $U_{(a,q)}^{0}$ is an additive subgroup of $\mathbb{F}_{q^{2}}.$\\
\item[(3)]  $U_{(a,q)}^{0}= \varepsilon^{i_{a}} U_{(1,q)}^{0}$, where $\varepsilon$ is a primitive element of $\mathbb{F}_{q^2}$ and $a=\varepsilon^{i_{a}(q-1)}.$\\
\item[(4)]  $A_{(a,q)}$ is also an additive subgroup of $\mathbb{F}_{q^{2}}.$\\
\item[(5)]  $U_{(a,q)}^{b}$'s are the additive cosets and $\mathbb{F}_{q^{2}}=\bigcup\limits_{b\in A_{(a,q)}}U_{(a,q)}^{b}$.  \\
\end{itemize}
\end{theorem}
$\mathbf{Proof.}$

 \begin{itemize}
\item[(1)] For any $b\in A_{(a,q)}$, then $b^{q}+a^{q}b=0$.\\

For any $\zeta\in U_{(a,q)}^{b}$, then $\zeta^{q}-a\zeta-b=0$. And then $$\zeta^{q^{2}}=(a\zeta+b)^{q}=a^{q}\zeta^{q}+b^{q}=a^{q}(a\zeta+b)+b^{q}=\zeta+a^{q}b+b^{q}=\zeta.$$
So $\zeta\in \mathbb{F}_{q^{2}}$ and then $U_{(a,q)}^{b}\subseteq \mathbb{F}_{q^{2}}$.\\

\item[(2)]  $U_{(a,q)}^{0}$ is an additive subgroup of $\mathbb{F}_{q^{2}}$.

By (1), we know that $U_{(a,q)}^{0}\subseteq \mathbb{F}_{q^{2}}$. And for any $\theta_{1},\theta_{2}\in U_{(a,q)}^{0}$, then

   $$(\theta_{1}-\theta_{2})^{q}-a(\theta_{1}-\theta_{2})=\theta_{1}^{q}-\theta_{2}^{q}-a(\theta_{1}-\theta_{2})  =(\theta_{1}^{q}-a\theta_{1})-(\theta_{2}^{q}-a\theta_{2})=0.$$

 So $\theta_{1}-\theta_{2}\in U_{(a,q)}^{0}$. Then $U_{(a,q)}^{0}$ is an additive subgroup of $\mathbb{F}_{q^{2}}$.\\

\item[(3)]$U_{(1,q)}^{0}=\{0\}\bigcup \{\alpha| \alpha^{q-1}=1\}$ and $U_{(a,q)}^{0}=\{0\}\bigcup \{\alpha| \alpha^{q-1}=a\}$. By the fact that $a^{q+1}=1$, let $\varepsilon$ be the primitive element of $\mathbb{F}_{q^2}$, then there exists some $i_{a}$, such that $a=(\varepsilon^{q-1})^{i_{a}}$. It can be seen that $U_{(a,q)}^{0}= \varepsilon^{i_{a}}U_{(1,q)}^{0}$.\\

 \item[(4)] Let $-a^{q}$ instead of $a$ in $U_{(a,q)}^{0}$ in (1), it is easy to see that $A_{(a,q)}=U_{(-a^{q},q)}^{0}$ is also a $q$ subgroup of $\mathbb{F}_{q^{2}}$.\\

 \item[(5)]  Let $\eta\in U_{(a,q)}^{b}$, then $\eta+U_{(a,q)}^{0}=U_{(a,q)}^{b}$.

 For any $\theta\in U_{(a,q)}^{0}$, then
 $$(\eta+\theta)^{q}-a(\eta+\theta)=\eta^{q}+\theta^{q}-a\eta-a\theta=(\eta^{q}-a\eta)+(\theta^{q}-a\theta)=b.$$
 So $\eta+U_{(a,q)}^{0}\subseteq U_{(a,q)}^{b}$ and $|\eta+U_{(a,q)}^{0}|=|U_{(a,q)}^{b}|$. Then $\eta+U_{(a,q)}^{0}=U_{(a,q)}^{b}$.

 Correspondingly, $\mathbb{F}_{q^{2}}=\bigcup\limits_{b\in A_{(a,q)}}U_{(a,q)}^{b}$.
\end{itemize}

\vspace{20pt}

\subsection{The First Class of Hermitian self-dual TGRS Codes}
\vspace{20pt}
In this subsection, we will present three constructions of the Hermitian self-dual TGRS code $TGRS_{n,k}[{\mathbf \alpha}, {\mathbf v}, A({\mathbf \eta})]$ over $\mathbb{F}_{q^2}$ with the following parameters:
\begin{itemize}
\item[(1)] The length $n|q-1,n=q,n=2\sqrt{q}$, respectively.
\item[(2)] The set of the evaluation points $\{\alpha_{1},\alpha_{2},\cdots,\alpha_{n}\}$ is a subset of a coset of additive subgroup of $\mathbb{F}_{q^{2}}$.
\item[(3)] $\frac{v_{i}^{q+1}}{u_{i}}=\lambda\in \mathbb{F}_{q^{2}}^{*}$, for $i=1,2,\cdots n$.
\end{itemize}

\vspace{20pt}

\begin{theorem}\label{f(x) and f(x+b)}
Let matrix $UA_{k}(b)=(a_{ij})$ be the $k\times k$ matrix with
\begin{equation}\label{matrix symbols 2}
a_{ij}=\left\{\begin{array}{cc}
C_{j-1}^{i-1}b^{j-i},&i\leq j,\\
0,&i>j,
\end{array}\right.
\end{equation}
where $C_{m}^{n}=\dfrac{m(m-1)\cdots(m-n+1)}{n(n-1)\cdots1}$ and $C_{m}^{0}=1$ for $m=0,1,\cdots,k-1$. Then $UA^{-1}_{k}(b)=UA_{k}(-b)$.
\end{theorem}

$\mathbf{Proof.}$

Consider the linear space $\mathbb{F}_{q}[x]_{k}$ over $\mathbb{F}_{q}$, where $\mathbb{F}_{q}[x]_{k}=\{a_{k-1}x^{k-1}+\cdots+a_{1}x+a_{0}|a_{i}\in \mathbb{F}_{q},i=0,1,\cdots,k-1\}$. Take $1,x,x^{2},\cdots,x^{k-1}$ and $1,x+b,(x+b)^{2},\cdots,(x+b)^{k-1}$ as two sets of basis. By the basic theory of linear algebra, we know that the transition matrix from the basis $1,x,x^{2},\cdots,x^{k-1}$ to $1,x+b,(x+b)^{2},\cdots,(x+b)^{k-1}$ is exactly $UA_{k}(b)$, i.e. $$(1,x+b,(x+b)^{2},\cdots,(x+b)^{k-1})=(1,x,x^{2},\cdots,x^{k-1})UA_{k}(b).$$ And then $(1,x,x^{2},\cdots,x^{k-1})=(1,x+b,(x+b)^{2},\cdots,(x+b)^{k-1})UA_{k}^{-1}(b)$. The $j$th column of $UA_{k}^{-1}(b)$ is exactly the coordinate of $x^{j-1}$ under the basis $1,x+b,\cdots,(x+b)^{k-1}$. Suppose $x^{j-1}=t_{0}+t_{1}(x+b)+\cdots+t_{k-1}(x+b)^{k-1}$, denote $x+b=y$, then it is converted to $(y-b)^{j-1}=t_{0}+t_{1}y+t_{2}y^{2}+\cdots+t_{k-1}y^{k-1}$. So $t_{0},t_{1},\cdots,t_{k-1}$ is also the coordinates of $(y-b)^{j-1}$ under the base $1,y,y^{2},\cdots,y^{k-1}$. By the above analysis, we can see that $t_{0},t_{1},\cdots,t_{k-1}$ is exactly the $j$th column of $UA_{k}(-b)$ for $j=1,2,\cdots,k$. So $UA_{k}^{-1}(b)=UA_{k}(-b)$.

\begin{theorem}\label{theorem of (+) twist}
Let $\mathbb{F}_{q^{2}}$ be a finite field with $q^{2}$ elements, where $q$ is a prime power. Let $n=2k$, $a\in \mathbb{F}_{q^{2}}$ satisfying $a^{q+1}=1$ and $b\in A_{(a,q)}$. $\{\alpha_{1},\alpha_{2},\cdots,\alpha_{n}\}\subseteq U_{(a,q)}^{b}$, and $\dfrac{v_{i}^{q+1}}{u_{i}}=\lambda\in \mathbb{F}_{q^{2}}^{*}$ for $i=1,2,\cdots,n$. Let $A({\mathbf \eta})\in \mathbb{F}_{q^{2}}^{k\times k}$ be a parameter matrix, then $TGRS_{n,k}({\mathbf \alpha}, {\mathbf v}, A({\mathbf \eta}))$ over $\mathbb{F}_{q^2}$ is Hermitian self-dual if and only if there exists an inverse matrix $P$ such that
\begin{equation}\label{equation for constructions 1-3}
[I_{k},A({\mathbf \eta})^{q}]=P[I_{k},-JA({\mathbf \eta})^{T}J]T({\mathbf\alpha})UA_{n}(-\dfrac{b}{a})^{T}\mathrm{diag}(1,\dfrac{1}{a},\dfrac{1}{a^{2}},\cdots,\dfrac{1}{a^{n-1}}),
\end{equation}
where $T({\mathbf\alpha}), A({\mathbf \eta})^{q}$ are denoted as in Eq. (\ref{matix symbols 1}) and $UA_{n}(-\dfrac{b}{a})$ is denoted as in Eq. (\ref{matrix symbols 2}).
\end{theorem}
$\mathbf{Proof.}$

Consider the following equation
$$(1,x+b,(x+b)^{2},\cdots,(x+b)^{n-1})=(1,x,x^{2},\cdots,x^{n-1})\left(\begin{array}{ccccc}
1&b&b^{2}&\cdots&b^{n-1}\\
0&1&C_{2}^{1}b&\cdots&C_{n-1}^{1}b^{n-2}\\
0&0&1&\cdots&C_{n-1}^{2}b^{n-3}\\
\vdots&\vdots&\vdots&\ddots&\vdots\\
0&0&0&\cdots&1
\end{array}
\right)$$
holds for any $x$. Because $\alpha_{1},\alpha_{2},\cdots,\alpha_{n}$ are the roots of $x^{q}-ax-b=0$, then $\alpha_{i}^{q}=a\alpha_{i}+b$ for $i=1,2,\cdots,n$. Then $$(1,\alpha_{i}^{q}, (\alpha_{i}^{q})^{2},\cdots, (\alpha_{i}^{q})^{n-1})=(1,\alpha_{i},\alpha_{i}^{2},\cdots,\alpha_{i}^{n-1})UA_{n}(\dfrac{b}{a})\mathrm{diag}(1,a,a^{2},\cdots,a^{n-1}).$$
So
$$\left(\begin{array}{cccc}
1&1&\cdots&1\\
\alpha_{1}^{q}&\alpha_{2}^{q}&\cdots&\alpha_{n}^{q}\\
\vdots&\vdots&\ddots&\vdots\\
(\alpha_{1}^{q})^{n-1}&(\alpha_{2}^{q})^{n-1}&\cdots&(\alpha_{n}^{q})^{n-1}
\end{array}
\right)=\mathrm{diag}(1,a,a^{2},\cdots,a^{n-1})UA_{n}(\dfrac{b}{a})^{T}\left(\begin{array}{cccc}
1&1&\cdots&1\\
\alpha_{1}&\alpha_{2}&\cdots&\alpha_{n}\\
\vdots&\vdots&\ddots&\vdots\\
\alpha_{1}^{n-1}&\alpha_{2}^{n-1}&\cdots&\alpha_{n}^{n-1}
\end{array}
\right).$$
 $\dfrac{v_{i}^{q+1}}{u_{i}}=\lambda$, so $\mathrm{diag}(\dfrac{v^{q+1}}{u})=\lambda I_{n}$.
Let  $G(x)=\prod\limits_{i=1}^{n}(x-\alpha_{i})=\sum\limits_{i=0}^{n-1}c_{i}x^{n-i}$, then by Theorem \ref{sufficient and necessary condtions for Hermitian self-dual matrix}, $C$ is Hermitian self-dual if and only if there exists an inverse matrix $P_{1}$ such that
$$[I_{k},A({\mathbf \eta})^{q}]V_{n}({\mathbf \alpha})^{q}\mathrm{diag}(\dfrac{v^{q+1}}{u})=P_{1}[I_{k},-JA({\mathbf \eta})^{T}J]T({\mathbf\alpha})V_{n}({\mathbf\alpha}),$$ which implies $$[I_{k},A({\mathbf \eta})^{q}]\mathrm{diag}(1,a,a^{2},\cdots,a^{n-1})UA_{n}(\dfrac{b}{a})^{T}=\dfrac{1}{\lambda}P_{1}[I_{k},-JA({\mathbf \eta})^{T}J]T({\mathbf\alpha}).$$

Let $\dfrac{1}{\lambda}P_{1}=P$, by Theorem \ref{f(x) and f(x+b)}, then

 $$ [I_{k},A({\mathbf \eta})^{q}]=P[I_{k},-JA({\mathbf \eta})^{T}J]T({\mathbf\alpha})UA_{n}(-\dfrac{b}{a})^{T}\mathrm{diag}(1,\dfrac{1}{a},\dfrac{1}{a^{2}},\cdots,\dfrac{1}{a^{n-1}}).$$

When the parameter matrix $A({\mathbf \eta})=0$, then the TGRS code $TGRS_{n,k}({\mathbf \alpha}, {\mathbf v}, A({\mathbf \eta}))$ degenerates into GRS code. And the existing Hermitian self-dual GRS codes can be viewed as special cases of  Theorem \ref{theorem of (+) twist} as follows:

{\bf Remark 2:} \begin{itemize}
\item[(1)]When $\lambda=1,a=1, A({\mathbf \eta})=0$, the Hermitian self-dual TGRS code constructed in Theorem \ref{theorem of (+) twist} is exactly the Hermitian self-dual GRS code constructed in \cite{37} Theorem 3.3.
\item[(2)]When $\lambda=\varepsilon^{m(n-1)}, a=1, A({\mathbf \eta})=0$, where $\varepsilon$ is a primitive element of $\mathbb{F}_{q^2}$ and $1\leq m\leq q$, the TGRS code constructed in Theorem \ref{theorem of (+) twist} is exactly the Hermitian self-dual GRS code constructed in \cite{37} Theorem 3.5.
\end{itemize}

\vspace{20pt}
{\bf Construction I}
\vspace{20pt}

\begin{corollary}\label{construction 1}
Let $\alpha_{1},\alpha_{2},\cdots,\alpha_{n}$ be the roots of $x^{n}-\delta=0$ where $q-1=nt$ and $\delta$ is a $t$-th primitive element in $\mathbb{F}_{q}^{*}$, and $\frac{v_{i}^{q+1}}{u_{i}}=\lambda\in \mathbb{F}_{q^{2}}^{*}$
for $i=1,2,\cdots,n$. Then $TGRS_{n,k}({\mathbf \alpha}, {\mathbf v}, A({\mathbf \eta}))$ over $\mathbb{F}_{q^2}$ is Hermitian self-dual if and only if $$A({\mathbf \eta})^{q}=-JA({\mathbf \eta})^{T}J.$$
\end{corollary}
$\mathbf{Proof.}$

Let $C$ be the TGRS code $TGRS_{n,k}[{\mathbf \alpha}, {\mathbf v}, A({\mathbf \eta})]$ over $\mathbb{F}_{q^2}$, where $\alpha_{1},\alpha_{2},\cdots,\alpha_{n}$ are the roots of $x^{n}-\delta=0$, $q-1=nt$, $\delta$ is a $t$-th primitive element in $\mathbb{F}_{q}^{*}$.
\begin{itemize}
\item[(1)]  Since $\alpha_{1},\alpha_{2},\cdots,\alpha_{n}$ are the roots of $x^{n}-\delta=0$, $q-1=nt$ and $\delta$ is a $t$-th primitive element. By Theorem \ref{Hermitian patition}, we have $$\{\alpha_{1},\alpha_{2},\cdots,\alpha_{n}\}\subseteq U_{(1,q)}^{0}.$$
\item[(2)] Let $G(x)=x^{n}-\delta$. Then the derivative $G'(x)=nx^{n-1}$. For each root $\alpha_{i}$, $G'(\alpha_{i})=n\alpha_{i}^{n-1}=n\dfrac{\alpha_{i}^{n}}{\alpha_{i}}=n\dfrac{\delta}{\alpha_{i}}\in \mathbb{F}_{q}$. Thus, there exists $v_{i}\in \mathbb{F}_{q^{2}}^{*}$, such that $$\dfrac{v_{i}^{q+1}}{u_{i}}=\lambda\in \mathbb{F}_{q^{2}}^{*},i=1,2,\cdots,n.$$
\end{itemize}
By (1)(2), it can be seen that $C$ is the TGRS code in the case where $a=1,b=0$ in Theorem \ref{theorem of (+) twist}. According to the selection of $\alpha_{i}$'s, the matrices $T({\mathbf\alpha})=UA_{n}(-\dfrac{b}{a})=\mathrm{diag}(1,\dfrac{1}{a},\cdots,\dfrac{1}{a})=I_{n}$ in  Eq. (\ref{equation for constructions 1-3}). Then, there exists an inverse matrix $P$ such that
$$[I_{k},A({\mathbf \eta})^{q}]=P[I_{k},-JA({\mathbf \eta})^{T}J]T({\mathbf\alpha})UA_{n}(-\dfrac{b}{a})^{T}\mathrm{diag}(1,\dfrac{1}{a},\dfrac{1}{a^{2}},\cdots,\dfrac{1}{a^{n-1}})$$
holds which is equivalent to $$[I_{k},A({\mathbf \eta})^{q}]=P[I_{k},-JA({\mathbf \eta})^{T}J].$$
Then, $P=I_{k}$. Therefore, $C$ is Hermitian self-dual if and only if $A({\mathbf \eta})^{q}=-JA({\mathbf \eta})^{T}J$.

{\bf Remark 3:} When $\delta=1, t=1,\lambda=1, A({\mathbf \eta})=0$, the TGRS code degenerates into the GRS code. And the TGRS code constructed in Corollary \ref{construction 1} is exactly the Hermitian self-dual GRS code constructed in \cite{38} Corollary 5.

\vspace{20pt}
{\bf Construction II}
\vspace{20pt}
\begin{corollary}\label{construction 2}
Let $\mathbb{F}_{q^{2}}$ be a finite field with $q^{2}$ elements, where $q=2^{s}$, $n=q$, $k=2^{s-1}$, and let $a\in\mathbb{F}_{q^{2}}$ satisfy $a^{q+1}=1$. Let $\{\alpha_{1},\alpha_{2},\cdots,\alpha_{n}\}$ be the roots of $x^{q}-ax=0$. Let $v_{1}=v_{2}=\cdots=v_{n}=\lambda\in \mathbb{F}_{q^{2}}^{*}$, and let $A({\mathbf \eta})\in\mathbb{F}_{q^{2}}^{k\times k}$ be an anti-diagonal matrix given by
 $$A({\mathbf \eta})=\left(\begin{array}{cccc}
 &&&0\\
 &&\eta_{k-1}&\\
 &\iddots&&\\
 \eta_{1}&&&
 \end{array}
 \right),$$ where $\eta_{i}^{q}+\dfrac{1}{a^{2i-1}}\eta_{i}=0$ for $i=1,2,\cdots,k-1$.
  Then $TGRS_{n,k}({\mathbf \alpha}, {\mathbf v}, A({\mathbf \eta}))$ over $\mathbb{F}_{q^2}$ is Hermitian self-dual.
\end{corollary}
$\mathbf{Proof.}$

\begin{itemize}
\item[(1)] Since $\alpha_{1},\alpha_{2},\cdots,\alpha_{n}$ are all roots of $x^{q}-ax-b=0$ for $b=0$. By Theorem \ref{Hermitian patition}, we have
    $$\{\alpha_{1},\alpha_{2},\cdots,\alpha_{n}\}=U_{(a,q)}^{0}.$$
\item[(2)] Let $G(x)=\prod\limits_{i=1}^{n}(x-\alpha_{i})=x^{q}-ax$. Then $G'(x)=qx^{q-1}-a$. So $G'(\alpha_{i})=-a=a$ for $i=1,2,\cdots,n$. Thus $$\dfrac{v_{i}^{q+1}}{u_{i}}=\dfrac{\lambda }{a}\in \mathbb{F}_{q^{2}}^{*}, i=1,2,\cdots,n.$$
\end{itemize}
By (1)(2), $C$ is the TGRS code in Theorem \ref{theorem of (+) twist} for $b=0$.

Since $\alpha_{1},\alpha_{2},\cdots,\alpha_{n}$ are all roots of $x^{q}-ax=0$, the matrices $UA_{n}(-\dfrac{b}{a})^{T}$ and $T({\mathbf\alpha})$ in Eq.(\ref{equation for constructions 1-3}) are as follows:  $UA_{n}(-\dfrac{b}{a})^{T}=I_{n}$ and $$T({\mathbf\alpha})=\left(\begin{array}{cccc}
 1&0&\cdots& 0\\
 0&1&\cdots&0\\
 \vdots&\vdots&\ddots&\vdots\\
 a&0&\cdots&1
 \end{array}
 \right).$$
 Then, consider the equation
$$P[I_{k},-JA({\mathbf \eta})^{T}J]T({\mathbf\alpha})UA_{n}(-\dfrac{b}{a})^{T}\mathrm{diag}(1,\dfrac{1}{a},\cdots,\dfrac{1}{a^{n-1}})
=P[I_{k},-JA({\mathbf \eta})^{T}J]T({\mathbf\alpha})\mathrm{diag}(1,\dfrac{1}{a},\cdots,\dfrac{1}{a^{n-1}})
=P[A_{1},A_{2}],$$
where $A_{1}=\begin{pmatrix}
1 & & & \\
& \dfrac{1}{a} & & \\
& & \ddots & \\
& & & \dfrac{1}{a^{k-1}}
\end{pmatrix}$ and $A_{2}=\begin{pmatrix}
 &  &  & 0 \\
 &  & -\dfrac{\eta_{k-1}}{a^{n-2}} &  \\
 & \iddots &  &  \\
-\dfrac{\eta_1}{a^{k}}&  &  &
\end{pmatrix}$.
 Let $P=\left(\begin{array}{cccc}
1&&&\\
&a&&\\
&&\ddots&\\
&&&a^{k-1}
\end{array}
\right)$. \\

It follows that $P[A_{1},A_{2}]=[I_{k},PA_{2}]$. Consider the matrix product: $$PA_{2}=\left(\begin{array}{ccccc}
&&&&0\\
&&&-\dfrac{\eta_{k-1}}{a^{n-3}}&\\
&&-\dfrac{\eta_{k-2}}{a^{n-5}}&&\\
&\iddots&&&\\
-\dfrac{\eta_{1}}{a}
\end{array}
\right)=\left(\begin{array}{ccccc}
&&&&0\\
&&&\eta_{k-1}^{q}&\\
&&\eta_{k-2}^{q}&&\\
&\iddots&&&\\
\eta_{1}^{q}
\end{array}
\right).$$
By Theorem \ref{theorem of (+) twist}, the $TGRS_{n,k}({\mathbf \alpha}, {\mathbf v}, A({\mathbf \eta}))$ is Hermitian self-dual.

{\bf Remark 4:}
\begin{itemize}
\item[(1)] When $a=1,\lambda =1,\eta_{2}=\eta_{3}=\cdots=\eta_{k-1}=0$, the Hermitian self-dual TGRS code constructed in Corollary \ref{construction 2} is exactly the Hermitian self-dual TGRS code constructed in \cite{12-1} Theorem 5.

\item[(2)] When $a=1,\lambda =1, A({\mathbf \eta})=0$, the Hermitian self-dual TGRS code constructed in Corollary \ref{construction 2} is exactly the Hermitian self-dual GRS code constructed in \cite{37} Theorem 3.3 for $a_{l}=0$.
\end{itemize}

\vspace{20pt}
{\bf Construction III}
\vspace{20pt}
\begin{corollary}\label{construction 3}
Let $q=p^{s}$ be a prime power, where $s$ is even. Let $q_{1}=p^{s/2}$, $n=2q_{1}=2k$. Suppose $\mathbb{F}_{q_{1}}\subset \mathbb{F}_{q}\subset \mathbb{F}_{q^{2}}$. Let $a,b\in \mathbb{F}_{q}$ satisfy $a^{q_{1}+1}=1,b\in A_{(a,q_{1})},b\neq0$, $c\in \mathbb{F}_{q^{2}}$ satisfy $c^{q+1}=1$. Let $\varepsilon$ be a primitive element of $\mathbb{F}_{q^2}$ and $c=\varepsilon^{i_{c}(q-1)}$ as defined in Theorem \ref{Hermitian patition} (3). Define the set $\{\alpha_{1},\alpha_{2},\cdots,\alpha_{n}\}=\varepsilon^{i_{c}}(U_{(a,q_{1})}^{0}\bigcup U_{(-a,q_{1})}^{-b})$. Let $\dfrac{v_{i}^{q+1}}{u_{i}}=\lambda \in\mathbb{F}_{q^{2}}^{*}$ for $1\leq i\leq n$. Let $A({\mathbf \eta})\in\mathbb{F}_{q^{2}}^{k\times k}$ be a diagonal matrix:
$$A({\mathbf \eta})=\left(\begin{array}{cccc}
\eta_{1}&&&\\
&\eta_{2}&&\\
&&\ddots&\\
&&&\eta_{k}
\end{array}
\right),
$$
where $\eta_{k}=0$, $\eta_{1}=0$, $1-b^{*}\eta_{k-i+1}\neq0$ and $\eta_{i}^{q}=-\dfrac{\eta_{k-i+1}}{c^{k}(1-b^{*}\eta_{k-i+1})}$ for $i=2,\cdots,k-1$, $b^{*}=\varepsilon^{i_{c}q_{1}}b$. Then $TGRS_{n,k}({\mathbf \alpha}, {\mathbf v}, A({\mathbf \eta}))$ is Hermitian sel-dual.
\end{corollary}
$\mathbf{Proof.}$

 Let $a,b\in \mathbb{F}_{q}$ satisfy $a^{q_{1}+1}=1, b\in A_{(a,q_{1})},b\neq0$, $U_{(a,q_{1})}^{b}$ be the set of the roots of $x^{q_{1}}-ax-b=0$. Then
\begin{itemize}
\item[(1)] $U_{(a,q_{1})}^{0}\bigcap U_{(-a,q_{1})}^{-b}=\emptyset.$

If $q=2^{s}$, then $a=-a$ and $U_{(-a,q_{1})}^{-b}=U_{(a,q_{1})}^{b}$ where $b\neq0$. By Theorem \ref{Hermitian patition} (5), $U_{(a,q_{1})}^{0}\bigcap U_{(-a,q_{1})}^{-b}=\emptyset$.

If $q$ is an odd prime power, suppose $\alpha\in U_{(a,q_{1})}^{0}\bigcap U_{(-a,q_{1})}^{-b}$, then we have

\begin{equation*}\label{construction 3 2n}
\left\{\begin{array}{cc}
\alpha^{q_{1}}-a\alpha=0 \ (A)\\
\alpha^{q_{1}}+a\alpha+b=0 \  (B)
\end{array}\right.
\end{equation*}
So $2a\alpha+b=0$. By substituting $\alpha=-\dfrac{b}{2a}$ into $(A)$, then we obtain $-\dfrac{b^{q_{1}}}{2a^{q_{1}}}+a\dfrac{b}{2a}=0$ which can be simplified into $-b^{q_{1}}+a^{q_{1}}b=0$. And because $a^{q_{1}+1}=1, b^{q_{1}}+a^{q_{1}}b=0$, then $b=0$ which contracts with $b\neq0$. So $U_{(a,q_{1})}^{0}\bigcap U_{(-a,q_{1})}^{-b}=\emptyset$.

\item[(2)] Let $\beta_{i}$ be the roots of $x^{q_{1}}-ax=0$ for $i=1,2,\cdots, q_{1}$ and $\beta_{j}$ be the roots of $x^{q_{1}}+ax+b=0$ for $j=q_{1}+1,q_{1}+2,\cdots,n$. Let $\alpha_{i}=\varepsilon^{i_{c}}\beta_{i}$, and denote $\varepsilon^{i_{c}}$ by $k_{c}$. Then
we have the polynomial:
\begin{align*}
G(x)&=\prod\limits_{i=1}^{n}(x-\alpha_{i})=\prod\limits_{i=1}^{n}(x-k_{c}\beta_{i})\\
&=k_{c}^{n}\prod\limits_{i=1}^{n}(\dfrac{x}{k_{c}}-\beta_{i})\\
&=k_{c}^{n}\left((\dfrac{x}{k_{c}})^{q_{1}}-a(\dfrac{x}{k_{c}})\right)\left((\dfrac{x}{k_{c}})^{q_{1}}+a(\dfrac{x}{k_{c}})+b\right)\\
&=(x^{q_{1}}-a^{*}x)(x^{q_{1}}+a^{*}x+b^{*}),
\end{align*}
where $a^{*}=ak_{c}^{q_{1}-1},b^{*}=k_{c}^{q_{1}}b$.\\

\item[(3)] For $1\leq i\leq n$, $G'(\alpha_{i})=-a^{*}(2a^{*}\alpha_{i}+b^{*})$ .\\

 Given $G(x)=\prod\limits_{i=1}^{n}(x-\alpha_{i})=(x^{q_{1}}-a^{*}x)(x^{q_{1}}+a^{*}x+b^{*})$, we compute its derivative: $G^{'}(x)=(q_{1}x^{q_{1}-1}-a^{*})(x^{q_{1}}+a^{*}x+b^{*})+(x^{q_{1}}-a^{*}x)(q_{1}x^{q_{1}-1}+a^{*})$. For any $\alpha_{i}$ that is a root of $G(x)$, we have $\alpha_{i}^{q_{1}}-a^{*}\alpha_{i}=0$ for $1\leq i\leq q_{1}$ and $\alpha_{i}^{q_{1}}+a^{*}\alpha_{i}+b^{*}=0$ for $q_{1}+1\leq i\leq n$.
 Substituting these into the derivative formula, we get: $$G^{'}(\alpha_{i})=-a^{*}(2a^{*}\alpha_{i}+b^{*}), 1\leq i\leq n.$$
 Additionally, consider the expression:
 \begin{align*}
 \dfrac{v_{i}^{q+1}}{u_{i}}&=v_{i}^{q+1}[-a^{*}(2a^{*}\alpha_{i}+b^{*})]\\
 &=-a^{*}[v_{i}^{q+1}(2a^{*}\alpha_{i}+b^{*})]\\
 &=-a^{*}[v_{i}^{q+1}(2ak_{c}^{q_{1}-1}k_{c}\beta_i+k_{c}^{q_{1}}b]\\
 &=-a^{*}k_{c}^{q_{1}}[v_{i}^{q+1}(2a\beta_{i}+b)].
 \end{align*}
 By Theorem \ref{Hermitian patition}, $\beta_{i},a, b\in\mathbb{F}_{q}$,
  there exists $v_{i}\in \mathbb{F}_{q^{2}}^{*}$, such that $v_{i}^{q+1}=(2\beta_{i}^{q_1}+b)^{-1}$. Let $-a^{*}k_{c}^{q_{1}}=\lambda$, then $\dfrac{v_{i}^{q+1}}{u_{i}}=\lambda$ for $1\leq i\leq n$.

 \item[(4)] $\alpha_{i}^{q}=c\alpha_{i}$, for $i=1,2,\cdots,n$.\\

By Theorem \ref{Hermitian patition}, $\mathbb{F}_{q}=\bigcup\limits_{b\in A_{(a,q_{1})}}U_{(a,q_{1})}^{b}$, In fact, $\mathbb{F}_{q}=U_{(1,q)}^{0}$. Thus, $U_{(a,q_{1})}^{0}\bigcup U_{(-a,q_{1})}^{-b}\subseteq U_{(1,q)}^{0}=\mathbb{F}_{q}$. Therefore,
$$\varepsilon^{i_{c}}(U_{(a,q_{1})}^{0}\bigcup U_{(-a,q_{1})}^{-b})\subseteq U_{(c,q)}^{0}.$$
It follows that $\alpha_{i}^{q}-c\alpha_{i}=0$ for $i=1,2,\cdots,n$ .
\end{itemize}
By (1)(2)(3)(4), we can see that $TGRS_{n,k}[{\mathbf \alpha}, {\mathbf v}, A({\mathbf \eta})]$ is the TGRS code in Theorem \ref{theorem of (+) twist} for $a=c,b=0$.
Moreover, the matrix $$UA_{n}(\dfrac{0}{c})^{T}=I_{n},
T({\mathbf \alpha})=\left(\begin{array}{cc}
I_{k}&0\\
A_{0}&I_{k}
\end{array}
\right)$$
in Eq.(\ref{equation for constructions 1-3}), where $I_{k}=\left(\begin{array}{cccc}
1&&&\\
&1&&\\
&&\ddots&\\
&&&1
\end{array}
\right)$ and $A_{0}=\left(\begin{array}{ccccc}
b^{*}&&&&\\
&b^{*}&&&\\
\vdots&&\ddots&&\\
-(a^{*})^{2}&\cdots&&\ddots&\\
-a^{*}b^{*}&-(a^{*})^{2}&&\cdots&b^{*}
\end{array}
\right).$ Let
 $$P= \begin{pmatrix}
\dfrac{1}{1-b^*\eta_k} & 0 & \cdots & 0 \\
0 & \dfrac{c}{1-b^*\eta_{k-1}} & \cdots & 0 \\
\vdots & \vdots & \ddots & \vdots \\
-\dfrac{(a^*)^2 \eta_2}{1-b^*\eta_k} & 0 & \cdots & 0 \\
-\dfrac{a^* b^* \eta_1}{1-b^*\eta_{k}} & -\dfrac{(a^*)^2 \eta_1}{1-b^*\eta_{k-1}} & \cdots & \dfrac{c^{k-1}}{1-b^*\eta_{1}}
\end{pmatrix},$$
then \begin{align*}
&P[I_{k},-JA({\mathbf \eta})^{T}J]T({\mathbf\alpha})UA_{n}(-\dfrac{b}{a})^{T}\mathrm{diag}(1,\dfrac{1}{c},\cdots,\dfrac{1}{c^{n-1}})\\
&=P[I_{k},-JA({\mathbf \eta})^{T}J]\left(\begin{array}{cc}
A_{1}&0\\
A_{2}&A_{3}
\end{array}
\right)\\
&=[P(A_{1}-JA({\mathbf \eta})^{T}JA_{2}),-PJA({\mathbf \eta})^{T}JA_{3}],
\end{align*}
 where
$A_{1}=\left(\begin{array}{cccc}
1&&&\\
&\dfrac{1}{c}&&\\
&&\ddots&\\
&&&\dfrac{1}{c^{k-1}}
\end{array}
\right)$,
$A_{2}=\left(\begin{array}{ccccc}
b^{*}&&&&\\
&\dfrac{b^{*}}{c}&&&\\
\vdots&&\ddots&&\\
-(a^{*})^{2}&\cdots&&\ddots&\\
-a^{*}b^{*}&\dfrac{-(a^{*})^{2}}{c}&&\cdots&\dfrac{b^{*}}{c^{k-1}}
\end{array}
\right)
$,

$A_{3}=\left(\begin{array}{cccc}
\dfrac{1}{c^{k}}&&&\\
&\dfrac{1}{c^{k-1}}&&\\
&&\ddots&\\
&&&\dfrac{1}{c^{n-1}}
\end{array}
\right)$.
We have:
\begin{align*}
&P(A_{1}-JA({\mathbf \eta})^{T}JA_{2})\\
&= \begin{pmatrix}
\dfrac{1}{1-b^*\eta_k} & 0 & \cdots & 0 \\
0 & \dfrac{c}{1-b^*\eta_{k-1}} & \cdots & 0 \\
\vdots & \vdots & \ddots & \vdots \\
-\dfrac{(a^*)^2 \eta_2}{1-b^*\eta_k} & 0 & \cdots & 0 \\
-\dfrac{a^* b^* \eta_1}{1-b^*\eta_{k}} & -\dfrac{(a^*)^2 \eta_1}{1-b^*\eta_{k-1}} & \cdots & \dfrac{c^{k-1}}{1-b^*\eta_{1}}
\end{pmatrix}\left(\begin{array}{cccc}
1-b^{*}\eta_{k}&&&\\
\vdots&\dfrac{1-b^{*}\eta_{k-1}}{c}&&\\
& \vdots&\ddots&\\
(a^{*})^{2}\eta_{2}&0&\cdots&\\
a^{*}b^{*}\eta_{1}&\dfrac{(a^{*})^{2}\eta_{1}}{c}&\cdots&\dfrac{1-b^{*}\eta_{1}}{c^{k-1}}
\end{array}
\right)\\
&=I_{n},
\end{align*}
and
$$-PJA({\mathbf \eta})^{T}JA_{3}=\begin{pmatrix}
\dfrac{-\eta_k}{(1-b^*\eta_k)c^k} & 0 & \cdots & 0 \\
0 & \dfrac{-c \eta_{k-1}}{(1-b^*\eta_{k-1})c^{k+1}} & \cdots & 0 \\
\vdots & \vdots & \ddots & \vdots \\
\dfrac{(a^*)^2\eta_2 \eta_k}{(1-b^*\eta_k)c^k} & 0 & \cdots & 0 \\
\dfrac{a^* b^* \eta_1 \eta_k}{(1-b^*\eta_k) c^k} & \dfrac{(a^*)^2\eta_1 \eta_{k-1}}{(1-b^*\eta_{k-1})c^{k+1}} & \cdots & \dfrac{-c^{k-1} \eta_1}{(1-b^*\eta_1)c^{n-1}}
\end{pmatrix}=\left(\begin{array}{cccc}
 \eta_{1}^{q}&&&\\
 &\eta_{2}^{q}&&\\
 &&\ddots&\\
 &&&\eta_{k}^{q}
 \end{array}
 \right),$$
for $\eta_{k}=0$ and $\eta_{1}=0$, $\eta_{i}^{q}=-\dfrac{\eta_{k-i+1}}{c^{k}(1-b^{*})\eta_{k-i+1}}$ for $i=2,\cdots,k-1$. Then,
$$[I_{k},A({\mathbf \eta})^{q}]=P[I_{k},-JA({\mathbf \eta})^{T}J]T({\mathbf\alpha})UA_{n}(-\dfrac{b}{a})^{T}\mathrm{diag}(1,\dfrac{1}{c},\dfrac{1}{c^{2}},\cdots,\dfrac{1}{c^{n-1}}).$$
By Theorem \ref{theorem of (+) twist}, $C$ is Hermitian self-dual.

Finally, in order to indicate the existence of the parameter matrix $A({\mathbf \eta})$ which satisfy the above supposed conditions, we need to analyze the existence of $\eta_{i}$'s such that
\begin{equation}\label{construction 3 eq}
 \eta_{i}^{q}=-\dfrac{\eta_{k-i+1}}{c^{k}(1-b^{*}\eta_{k-i+1})}
\end{equation}
 hold. From the expression, we can see that $ \eta_{i}^{q}=-\dfrac{\eta_{k-i+1}}{c^{k}(1-b^{*}\eta_{k-i+1})} $ and $\eta_{k-i+1}^{q}=-\dfrac{\eta_{i}}{c^{k}(1-b^{*}\eta_{i})}$ need to hold simultaneously.

For any fixed $\eta_{i}\in \mathbb{F}_{q^{2}}$, let $\eta_{k-i+1}^{q}=-\dfrac{\eta_{i}}{c^{k}(1-b^{*}\eta_{i})}$ and then $\eta_{k-i+1}=-\dfrac{\eta_{i}^{q}}{c^{kq}\left(1-(b^{*})^{q}\eta_{i}^{q}\right)}$.
And
\begin{align*}
-\dfrac{\eta_{k-i+1}}{c^{k}(1-b^{*}\eta_{k-i+1})}&=-\dfrac{-\dfrac{\eta_{i}^{q}}{c^{kq}\left(1-(b^{*})^{q}\eta_{i}^{q}\right)}}{c^{k}\left(1+b^{*}\dfrac{\eta_{i}^{q}}{c^{kq}\left(1-(b^{*})^{q}\eta_{i}^{q}\right)}\right)}\\
&=\dfrac{\eta_{i}^{q}}{c^{k}\left( c^{kq}(1-(b^{*})^{q}\eta_{i}^{q})+b^{*}\eta_{i}^{q}\right)}\\
&=\dfrac{\eta_{i}^{q}}{c^{k(q+1)}\left(1-(b^{*})^{q}\eta_{i}^{q}\right)+c^{k}b^{*}\eta_{i}^{q}}\\
&=\dfrac{\eta_{i}^{q}}{(1-(b^{*})^{q}\eta_{i}^{q})+c^{k}b^{*}\eta_{i}^{q}}\\
&=\dfrac{\eta_{i}^{q}}{1+(c^{k}b^{*}-(b^{*})^{q})\eta_{i}^{q}}.
\end{align*}
By the fact that $b^{*}=k_{c}^{q_{1}}b$, $k_{c}=\varepsilon^{i_{c}}$ and $\varepsilon^{i_{c}(q-1)}=c$. So
$$c^{k}b^{*}-(b^{*})^{q}=b^{*}(c^{k}-(b^{*})^{q-1})=b^{*}(c^{k}-\varepsilon^{i_{c}q_{1}(q-1)}b^{q-1})=b^{*}(c^{k}-c^{q_{1}})=0.$$ Then $$-\dfrac{\eta_{k-i+1}}{c^{k}(1-b^{*}\eta_{k-i+1})}=\eta_{i}^{q}.$$ So when $i\neq k-i+1$, i.e. $k$ is even, such pairs always exist.

So when $q=2^{s}$, the parameters $\eta_{i}$'s such that Eq. (\ref{construction 3 eq}) hold always exist.

When $q$ is an odd prime, for $i\neq \frac{k+1}{2}$, by the analysis above, the pairs $\eta_{i}, \eta_{k-i+1}$ such that
\begin{center}
$ \eta_{i}^{q}=-\dfrac{\eta_{k-i+1}}{c^{k}(1-b^{*}\eta_{k-i+1})} $ and $\eta_{k-i+1}^{q}=-\dfrac{\eta_{i}}{c^{k}(1-b^{*}\eta_{i})}$
\end{center}
hold simultaneously always exist. And for $i_{0}=\frac{k+1}{2}$, then $\eta_{i_{0}}=\eta_{k-i_{0}+1}$. So Eq. (\ref{construction 3 eq}) turns to $$\eta_{i_{0}}^{q}=-\dfrac{\eta_{i_{0}}}{c^{k}(1-b^{*}\eta_{i_{0}})}.$$
Then $\eta_{i_{0}}=0$ or $\eta_{i_{0}}$ is a solution of equation
\begin{equation}\label{construction 3  eq parameter analysis}
c^{k}b^{*}x^{q}-c^{k}x^{q-1}-1=0.
\end{equation}
The reciprocal polynomial of Eq. (\ref{construction 3  eq parameter analysis}) is
\begin{equation}\label{construction 3  eq2 parameter analysis}
 x^{q}+c^{k}x-c^{k}b^{*}=0.
 \end{equation}
Because $(-c^{k})^{q+1}=1$  and
\begin{align*}
(c^{k}b^{*})^{q}+(-c^{k})^{q}c^{k}b^{*}&=c^{kq}(b^{*})^{q}-c^{k(q+1)}b^{*}\\
&=c^{kq}(b^{*})^{q}-b^{*}\\
&=b^{*}(c^{kq}(b^{*})^{q-1}-1)\\
&=b^{*}(c^{kq}(\varepsilon^{i_{c}q_{1}})^{q-1}-1)\\
&=b^{*}(c^{kq}c^{k}-1)\\
&=b^{*}(c^{k(q+1)}-1)\\
&=0.
\end{align*}
By Theorem \ref{Hermitian patition}, the $q$ solutions of Eq. (\ref{construction 3  eq2 parameter analysis}) are all in $\mathbb{F}_{q^{2}}$. So such $\eta_{i_{0}}$ also exist.

Based on the above analysis, there are many $\eta_{i}$'s in $\mathbb{F}_{q^{2}}$ such that $\eta_{i}^{q}=-\dfrac{\eta_{k-i+1}}{c^{k}(1-b^{*})\eta_{k-i+1}}$ for $i=2,\cdots,k-1$ hold.

\vspace{20pt}
\subsection{The Second Class of Hermitian Self-dual TGRS Codes}
\vspace{20pt}
In this subsection, we will consider the Hermitian self-dual TGRS codes with the following parameters:

\begin{itemize}
\item[(1)] The length $n=q+1$.
\item[(2)] The set of the evaluation points $\{\alpha_{1},\alpha_{2},\cdots,\alpha_{n}\}$ is a subset of a coset of multiplicative subgroup of $\mathbb{F}_{q^{2}}^{*}$.
\item[(3)] $\frac{v_{i}^{q+1}}{u_{i}}\neq \frac{v_{j}^{q+1}}{u_{j}}$, for $1\leq i\neq j\leq n$.
\end{itemize}

\vspace{20pt}
{\bf Construction IV}
\vspace{20pt}
\begin{theorem}\label{construction 4}
 Let $\mathbb{F}_{q^{2}}$ be a finite field with $q^{2}$ elements, where $q$ is an odd prime power. Let $b\in \mathbb{F}_{q}^{*}$ and $\{\alpha_{i}|i=1,2,\cdots,n\}$ be the roots of $x^{n}-b$, where $n=q+1=2k$. Let $v_{i}^{q+1}=\lambda\in \mathbb{F}_{q}^{*},i=1,2,\cdots,n$ and  $A({\mathbf \eta})\in\mathbb{F}_{q^{2}}^{k\times k}$ be a diagonal matrix:
$$A({\mathbf \eta})=\left(\begin{array}{cccc}
\eta_{1}&&&\\
&\eta_{2}&&\\
&&\ddots&\\
&&&\eta_{k}
\end{array}
\right),
$$
where $\eta_{i}=0$ or $\eta_{i}\neq0$ and $-\dfrac{1}{\eta_{k-i+1}b^{2i+1}}=\eta_{i}^{q}$, $i=1,2,\cdots,k$. Then the TGRS code $TGRS_{k}({\mathbf\alpha},A({\mathbf \eta}))$ is Hermitian self-dual.
\end{theorem}
$\mathbf{Proof.}$

\begin{itemize}
\item[(1)]
Let $G(x)=\prod\limits_{i=1}^{n}(x-\alpha_{i})=x^{q+1}-b$. Then, note that $T(\alpha)=I_{n}.$ Compute the deritive
$G'(x)=(q+1)x^{q}$. For each root $\alpha_{i}$, we have $G'(\alpha_{i})=\alpha_{i}^{q}=\dfrac{b}{\alpha_{i}}$. Thus, we get $$u_{i}=\dfrac{\alpha_{i}}{b}, i=1,2,\cdots,n.$$

\item[(2)]
Let $\alpha_{i}$ be a root of $x^{q+1}=b$. Then $\alpha_{i}^{q+1}=b$, and it follows that $\alpha_{i}^{q}=\dfrac{b}{\alpha_{i}}$, for $i=1,2,\cdots,n$. Then
\begin{align*}
V_n({\mathbf\alpha})^q
&= \begin{pmatrix}
1 & 1 & \cdots & 1 \\
\alpha_1^q & \alpha_2^q & \cdots & \alpha_n^q \\
\vdots & \vdots & \ddots & \vdots \\
(\alpha_1^q)^{n - 1} & (\alpha_2^q)^{n - 1} & \cdots & (\alpha_n^q)^{n - 1}
\end{pmatrix}\\
&= \begin{pmatrix}
1 & 1 & \cdots & 1 \\
\dfrac{b}{\alpha_1} & \dfrac{b}{\alpha_2} & \cdots & \dfrac{b}{\alpha_n} \\
\vdots & \vdots & \ddots & \vdots \\
\bigl(\dfrac{b}{\alpha_1}\bigr)^{n - 1} & \bigl(\dfrac{b}{\alpha_2}\bigr)^{n - 1} & \cdots & \bigl(\dfrac{b}{\alpha_n}\bigr)^{n - 1}
\end{pmatrix}\\
& = \begin{pmatrix}
1 \\
& b \\
& & \ddots \\
& & & b^{n-1}
\end{pmatrix} V_n\Bigl(\dfrac{1}{{\mathbf\alpha}}\Bigr)\\
&=\mathrm{diag}(1,b,\cdots,b^{n-1})JV_{n}({\mathbf\alpha})\mathrm{diag}(\dfrac{1}{\alpha_{1}^{n-1}},\dfrac{1}{\alpha_{2}^{n-1}},\cdots,\dfrac{1}{\alpha_{n}^{n-1}}).
\end{align*}

\item[(3)]
Then, consider the equation $$[I_{k},A({\mathbf \eta})^{q}]V_{n}({\mathbf \alpha})^{q}\mathrm{diag}(\dfrac{{\mathbf v}^{q+1}}{\mathbf u})=P[I_{k},-JA({\mathbf \eta})^{T}J]T({\mathbf\alpha})V_{n}({\mathbf \alpha}).$$
This transforms to $$[I_{k},A({\mathbf \eta})^{q}]\mathrm{diag}(1,b,\cdots,b^{n-1})JV_{n}({\mathbf\alpha})\mathrm{diag}(\dfrac{1}{\alpha_{1}^{n-1}},
\dfrac{1}{\alpha_{2}^{n-1}},\cdots,\dfrac{1}{\alpha_{n}^{n-1}})
\mathrm{diag}(\dfrac{{\mathbf v}^{q+1}}{\mathbf u})=P[I_{k},-JA({\mathbf \eta})^{T}J]V_{n}({\mathbf \alpha}),$$
which is equivalent to: $$[I_{k},A({\mathbf \eta})^{q}]\mathrm{diag}(1,b,\cdots,b^{n-1})JV_{n}({\mathbf\alpha})\mathrm{diag}(\dfrac{{\mathbf v}^{q+1}}{{\mathbf u}{\mathbf\alpha}^{n-1}})=P[I_{k},-JA({\mathbf \eta})^{T}J]V_{n}({\mathbf\alpha}).$$

Note that $\dfrac{v_{i}^{q+1}}{u_{i}\alpha_{i}^{n-1}}=v_{i}^{q+1}=\lambda$ for $i=1,2,\cdots,n$. Then the above equation simplifies to:

$$[I_{k},A({\mathbf \eta})^{q}]\mathrm{diag}(1,b,\cdots,b^{n-1})JV_{n}({\mathbf\alpha})=P[I_{k},-JA({\mathbf \eta})^{T}J]V_{n}({\mathbf \alpha}),$$
which is equivalent to: $$[I_{k},A({\mathbf \eta})^{q}]\mathrm{diag}(1,b,\cdots,b^{n-1})J=P[I_{k},-JA({\mathbf \eta})^{T}J].$$ In other words:
$$[I_{k},A({\mathbf \eta})^{q}]=P[I_{k},-JA({\mathbf \eta})^{T}J]J\mathrm{diag}(1,\dfrac{1}{b},\cdots,\dfrac{1}{b^{n-1}}).$$

Define $[I_k, -JA({\mathbf \eta})^T J]J \, \mathrm{diag}\left(1, \dfrac{1}{b}, \dots, \dfrac{1}{b^{n-1}}\right)$ as
$[A_1,A_2],$

where
$
A_1 = \begin{pmatrix}
 &  &  & -\dfrac{\eta_k}{b^{k-1}} \\
 &  & \iddots &  \\
 & -\dfrac{\eta_{2}}{b} &  &  \\
-\eta_1 &  &  &
\end{pmatrix},
$
$
A_2 = \begin{pmatrix}
 &  &  & \dfrac{1}{b^{n-1}} \\
 &  & \iddots &  \\
 & \dfrac{1}{b^{k+1}} &  &  \\
\dfrac{1}{b^{k}}&  &  &
\end{pmatrix}.
$
\end{itemize}

If $\eta_{i}=0$, $i=1,2,\cdots,k$, let $P=\left(\begin{array}{cccc}
1&&&\\
&b&&\\
&&\ddots&\\
&&&b^{k-1}
\end{array}
\right)$. We have $[I_{k},A({\mathbf \eta})^{q}]=P[A_{1},A_{2}].$

If $\eta_{i}\neq0$, $i=1,2,\cdots,k$,
let
 $P=\left(\begin{array}{cccc}
&&&-\dfrac{b^{k-1}}{\eta_{k}}\\
&&-\dfrac{b^{k-2}}{\eta_{k-1}}&\\
&\iddots&&\\
-\dfrac{1}{\eta_{1}}&&&
\end{array}
\right).
$ Then
$PA_{1}=I_{k}$. Given that  $-\dfrac{1}{\eta_{k-i+1}b^{2i+1}}=\eta_{i}^{q}$, $i=1,2,\cdots,n$, consider\\

$$PA_{2}=\left(\begin{array}{cccc}
-\dfrac{1}{b\eta_{k}}&&&\\
&-\dfrac{1}{b^{3}\eta_{k-1}}&&\\
&&\ddots&\\
&&&-\dfrac{1}{b^{n-1}\eta_{1}}
\end{array}
\right)=\left(\begin{array}{cccc}
\eta_{1}^{q}&&&\\
&\eta_{2}^{q}&&\\
&&\ddots&\\
&&&\eta_{2}^{q}
\end{array}
\right),$$
that is $PA_{2}=A({\mathbf \eta})^{q}$.
 By Theorem \ref{sufficient and necessary condtions for Hermitian self-dual matrix}, $C$ is a Hermitian self-dual TGRS code.

{\bf Remark 5:}

When $b=1,A({\mathbf \eta})=0$, the TGRS code degenerates into the GRS code. And the TGRS code constructed in Theorem \ref{construction 4} has a relationship with the Hermitian self-dual GRS code constructed in \cite{38} Corollary 8.

\section{Hermitian self-dual MDS A-TGRS codes}
For the convenience of the following description, we give some notations. Let $I$ be any $k$-subset of $\{1,2,\cdots,n\}$. We list the following series of symbols:
\begin{equation}\label{12}
G(x)=\prod\limits_{ i\in I}(x-\alpha_{i})=c_{0}x^{k}+c_{1}x^{k-1}+\cdots+c_{k-1}x+c_{k},
\end{equation}
\begin{equation}\label{8}
A_{I}=\left(\begin{array}{cccccc}
0 & 1 & & &\\
0 & 0 & 1 &  &\\
0&  0& 0&\ddots &\\
  \vdots & \vdots  & \vdots  &  \ddots&    1   &\\
 -c_{k}&-c_{k-1}& -c_{k-2}&\cdots&-c_{1}\\
\end{array}
\right),
\end{equation}
\begin{equation}\label{9}
d_{j}=c_{k-j}, \ a_{m,t}^{l}=\sum\limits_{
 i+j=l, 1\leq i\leq n-k, 0\leq j\leq t-1}\eta_{m,i}d_{j},
\end{equation}
\begin{equation}\label{10}
F_{m,t}(x)=\sum\limits_{l=t}^{n-k+t-1}a_{m,t}^{l}x^{l-t},\  g_{m,t}=-\boldsymbol{\gamma} F_{m,t}(A_{I})\boldsymbol{\gamma^{T}},
\end{equation}
{\footnotesize\begin{equation}\label{11}
M(n,k,\boldsymbol{{\mathbf\alpha}},A({\mathbf \eta}),I)=\left|\begin{array}{cccc}
 1+g_{0,1}&g_{0,2}&\cdots&g_{0,k}\\
 g_{1,1}&1+g_{1,2}&\cdots&g_{1,k}\\
 \vdots&\vdots&\ddots&\vdots\\
 g_{k-1,1}&g_{k-1,2}&\cdots&1+g_{k-1,k}
 \end{array}
 \right|,
\end{equation}}

 where $\boldsymbol{\gamma}=(0,\cdots,0,1)$, $j=0,1,\cdots,k$, $m=0,1,\cdots k-1$, $t=1,2,\cdots,k$, $l=1,2,\cdots,n-1$.

\begin{lemma}( Theorem 6 in \cite{29}) \label{TGRS MDS sufficient and necessary condtions}
Suppose that $3\leq k<n$ and $\alpha_{1},\alpha_{2},\cdots,\alpha_{n}\in \mathbb{F}_{q}$ are distinct elements. Let $\Omega$ be the set of $A({\mathbf \eta})$ such that
\begin{equation}\label{parameter equation}
M(n,k,{\mathbf\alpha},A({\mathbf \eta}),I)\neq0
\end{equation}
for each $k$-subset $I$ of $\{1,2,\cdots,n\}$. Then the code $C$ constructed by (\ref{generator matrix 1}) is MDS if and only if $$A({\mathbf \eta})\in\Omega.$$
\end{lemma}
From Theorem \ref{sufficient and necessary condtions for Hermitian self-dual matrix} and Lemma \ref{TGRS MDS sufficient and necessary condtions}, we can give the sufficient and necessary condtions for MDS Hermitian self-dual codes.

\begin{theorem}\label{sufficient and necessary condtions for MDS Hermitian self-dual}
Let $\mathbb{F}_{q^{2}}$ be a finite field with $q^{2}$ elements, where $q$ is a prime power. Let $\alpha_{1},\alpha_{2},\cdots,\alpha_{n}$ be different elements in $\mathbb{F}_{q^{2}}$ and $n=2k$. Let $A({\mathbf \eta})\in \mathbb{F}_{q^{2}}^{k\times k}$ be a parameter matrix. Then the TGRS code $TGRS_{n,k}({\mathbf \alpha},{\mathbf v}, A({\mathbf \eta}))$ is MDS Hermitian self-dual if and only if
\begin{itemize}
\item[(i)] There exists an inverse matrix $P$ such that
$[I_{k},A({\mathbf \eta})^{q}]V_{n}({\mathbf \alpha})^{q}\mathrm{diag}(\dfrac{{\mathbf v}^{q+1}}{\mathbf u})=P[I_{k},-JA({\mathbf \eta})J]T({\mathbf \alpha})V_{n}({\mathbf \alpha})$;\\
\item[(ii)] For each $k$-subset $I$ of $\{1,2,\cdots,n\}$, $M(n,k,{\mathbf\alpha},A({\mathbf \eta}),I)\neq0$.\\
\end{itemize}
\end{theorem}

{\bf Construction V}\\
Let $\mathbb{F}_{q^{2}}$ be a finite field of $q^{2}$ elements, where $q$ is an odd prime power. Let $\mathbb{F}_{q}$ be a proper subfield of $\mathbb{F}_{q^{2}}$. Let $\alpha_{1},\alpha_{2},\cdots,\alpha_{n}$ be the roots of $x^{n}-\delta$, where $\delta$ is a $t$-th primitive element in $\mathbb{F}_{q}^{*}$ and $q-1=nt$. Let $\frac{v_{i}^{q+1}}{u_{i}}=\lambda\in \mathbb{F}_{q^{2}}^{*}$ for $i=1,2,\cdots,n$.
Consider the parameter matrix
 $$A({\mathbf \eta})=\left(\begin{array}{cccc}
\eta&\eta&\cdots&\eta\\
\eta&\eta&\cdots&\eta\\
\vdots&\vdots&\ddots&\vdots\\
\eta&\eta&\cdots&\eta\\
\end{array}
\right),$$
where $\eta^{q}=-\eta$. By the proof of Theorem \ref{construction 1}, there exists an inverse matrix $P=I_{k}$ such that \\

$(i)$ $[I_{k},A({\mathbf \eta})^{q}]V_{n}({\mathbf \alpha})^{q}\mathrm{diag}(\dfrac{{\mathbf v}^{q+1}}{\mathbf u})=P[I_{k},-JA({\mathbf \eta})J]T({\mathbf \alpha})V_{n}({\mathbf \alpha})$ holds.\\

$(ii)$  When $\eta^{q}=-\eta$, then $\eta\in \mathbb{F}_{q^2}  \backslash \mathbb{F}_{q}$. And $\alpha_{1},\alpha_{2},\cdots,\alpha_{n}\in \mathbb{F}_{q}$.

For any $k$-subset $I$ of $\{1,2,\cdots,n\}$, the diagram
 \begin{align*}
M(n,k,\boldsymbol{{\mathbf\alpha}},A({\mathbf \eta}),I)&=\left|\begin{array}{cccc}
 1+g_{0,1}&g_{0,2}&\cdots&g_{0,k}\\
 g_{1,1}&1+g_{1,2}&\cdots&g_{1,k}\\
 \vdots&\vdots&\ddots&\vdots\\
 g_{k-1,1}&g_{k-1,2}&\cdots&1+g_{k-1,k}
 \end{array}
 \right|\\
 &=\left|I_{k}+\left(\begin{array}{c}
 \eta\\
 \eta\\
 \vdots\\
 \eta
 \end{array}
 \right)(a_{0,1},a_{0,2},\cdots,a_{0,k})\right|\\
 \\
 &=\left|1+(a_{0,1},a_{0,2},\cdots,a_{0,k})\left(\begin{array}{c}
 \eta\\
 \eta\\
 \vdots\\
 \eta
 \end{array}
 \right)\right|\\
 &=1+\sum\limits_{t=1}^{k}a_{0,t}\eta=1-\sum\limits_{t=1}^{k}\eta\sum_{j=1}^{t}d_{t-j}\sum_{l=k}^{n-1}w_{l-j},\\
 \end{align*}
 where $a_{0,t}\eta=g_{0,t}$ for $t=1,\cdots,k$. By Eqs. (\ref{12})(\ref{8})(\ref{9})(\ref{10})(\ref{11}), since $\boldsymbol{{\mathbf\alpha}}\in \mathbb{F}_{q}^{n}$, we have $\sum\limits_{t=1}^{k}\sum_{j=1}^{t}d_{t-j}\sum_{l=k}^{n-1}w_{l-j}\in \mathbb{F}_{q}$. Given $\eta\in \mathbb{F}_{q^2}  \backslash \mathbb{F}_{q}$, it follows that $M(n,k,\boldsymbol{{\mathbf\alpha}},A({\mathbf \eta}),I)\neq0$ for any $k$-subset $I$.

 By Theorem \ref{sufficient and necessary condtions for MDS Hermitian self-dual}, $C$ is an MDS self-dual TGRS code.

\section{Conclusion}
Our work mainly consists of three parts. First, we present the sufficient and necessary condition under which an A-TGRS code is Hermitian self-dual. Second, we propose four constructions which not only include all the existing Hermitian self-dual TGRS codes but also produce many new ones. Third, we establish the sufficient and necessary condition for an A-TGRS code to be a Hermitian self-dual MDS code. Additionally, we provide a construction method for Hermitian self-dual MDS TGRS codes.

\end{document}